\documentclass[aps,epsf]{iopart}
\usepackage{graphics}
\usepackage{graphicx}
\usepackage{epsfig}

\begin{document}

\title{A cosmological model with complex scalar field}
\author{Xiang-hua Zhai, Yi-bin Zhao}

\address{Shanghai United Center for Astrophysics(SUCA), Shanghai Normal
University, 100 Guilin Road, Shanghai 200234,China}

\ead{kychz@shnu.edu.cn} \vspace{1cm}
\begin{abstract}
In this paper, we wish to point out the possibility of using a
complex scalar field to account for the inflationary stage and the
current acceleration. By the analysis of the dynamical system and
numerical work, we show that the amplitude of the complex scalar
field plays the role of the inflaton whereas the phase is the
quintessence field. The numerical solutions describe heteroclinic
orbits, which interpolate between an unstable critical point and a
late-time de Sitter attractor. Therefore, this model is more
natural to explain the two stages of acceleration.
\end{abstract}


\maketitle

\section{Introduction}
Astronomical observation on the cosmic microwave background(CMB)
anisotropy \cite{CMB}, supernova type Ia(SNIa) \cite{SNIa} and
SLOAN Digital Sky Survey(SDSS) \cite{SDSS} converge on that our
Universe is spatially flat, with about $70\%$ of the total density
resulting from dark energy that has an equation of state $w<-1/3$
and drives the accelerating expansion of the Universe which began
at a redshift of order one-half. The origin of the dark energy
remains elusive from the point of view of general relativity and
standard particle physics. To represent dark energy, several
candidates have been suggested and confronted with observation: a
cosmological constant, quintessence with a single field
\cite{SingleField} or with $N$ coupled field \cite{NCoupledField},
a phantom field with a canonical \cite{Caldwell} or Born-Infield
type Lagrangian \cite{BIL}, $k$-essence \cite{k-essence} and the
generalized Chaplygin gas(GCG) \cite{GCG}. Among these models, the
most typical ones are cosmological constant and quintessence which
has caught much attention ever since its invention.

The idea of inflation is legitimately regarded as an great
advancement of modern cosmology: it solves the horizon, flatness
and monopole problem, and it provides a mechanism for the
generation of density perturbations needed to seed the formation
of structures in the universe \cite{Kolb&Turner}. In standard
inflationary models \cite{Liddle&Lyth}, the physics lies in the
inflation potential. The underlying dynamics is simply that of a
single scalar field rolling in its potential. This scenario is
generically referred to as chaotic inflation in reference to its
choice of initial conditions. This picture is widely favored
because of its simplicity and has received by far the most
attention to date. The properties of inflationary models are also
tightly constraint by the recent result from the observation. The
standard inflationary $\Lambda$CDM model provides a good fit to
the observed cosmic microwave background (CMB) anisotropy. Peebles
and Vilenkin proposed and quantitatively analyzed the intriguing
idea \cite{QuintInf} that a substantial fraction of the present
cosmic energy density could reside in the vacuum potential energy
of the scalar field responsible for inflation (quintessential
inflation). After that, there were some models to be presented in
succession.

In this paper, we wish to point out the possibility of using a
complex scalar field to account for the inflationary stage and the
current acceleration. We put the emphasis on the study of the
dynamics. According to the phase space analysis and numerical
calculation, we show that how the amplitude of complex field plays
the role of the inflaton whereas the phase is the quintessence
field. There are stable and unstable critical points for our
model. In our case, one of the unstable critical point is
corresponding to the inflationary stage at very first moments of
the Universe and the stable critical point is corresponding to the
second stage of accelerated expansion began at a redshift $z \sim
1.5$ and is still operative, which alleviates the fine tuning
problem. The equation of state $w$ varies with the cosmic
evolution and approaches towards -1 asymptotically showing the
existence of a cosmological constant at late times. The
heteroclinic orbits \cite{heteroclinic} connect unstable and
stable critical points. The numerical calculation shows that
brought the Universe back to the usual Friedman-Robertson-Walker
expansion, then the second stage of accelerated expansion began at
$z \sim 1.5$.

\section{The basics}
Since current observations favour a flat Universe, we will work in
the spatially flat Robertson-Walker metric,

\begin{equation}\label{metric}
ds^2=dt^2-a^2(t)d\textbf{x}^2,
\end{equation}
the Lagrangian density for the spatially homogeneous complex
scalar field $\Phi $ is
\begin{equation}\label{Lagrangian}
\mathcal{L}_{\Phi } = \frac{1}{2}g^{\mu \nu}\left( \partial
_{\mu}\Phi^{*} \right) \left( \partial _{\nu}\Phi \right)- V(|
\Phi|)-U(\theta),
\end{equation}
when consider the presence of barotropic fluid, the action for the
model is
\begin{equation}\label{action}
S = \int d^{4}x\sqrt{g} \left( -\frac{1}{2 \kappa^2}\mathcal{R} -
\rho _{\gamma} + \mathcal{L}_{\Phi} \right),
\end{equation}
where $g$ is the determinant of the metric tensor $g_{\mu\nu}$,
$\kappa^2=8\pi G$, $\mathcal{R}$ is the Ricci scalar, and
$\rho_{\gamma}$ is the density of the fluid with a barotropic
equation of state $p_{\gamma}=(\gamma-1)\rho_{\gamma}$, where
$0\leq \gamma\leq2$ is a constant that relates to the equation of
state by $w=\gamma-1$.

Writing the complex field $\Phi$ as (see \cite{complexscalar})
\begin{equation}\label{definition}
\Phi (t) = \frac{1}{\sqrt{2}} \phi (t) e^{i \theta (t)/f},
\end{equation}
Substituting (\ref{definition}) into (\ref{Lagrangian}) and
varying the action, one can obtain the Einstein equations and the
equations of motion for the scalar field as

\begin{eqnarray}
\label{H2}
H^2 = \frac{\kappa^2}{3} \left[ \rho _{\gamma} +
\frac{1}{2} \left(  \dot{\phi}^{2} + \frac{ \phi ^{2}}{f^2} \dot{\theta}^{2} \right )
 + V(\phi )+U(\theta)  \right],\\
\label{dotH} \dot H = -\frac{\kappa^2}{2}
\left[\rho_{\gamma}+p_\gamma + \dot{\phi}^{2}+ \frac{
\phi^{2}}{f^2} \dot{\theta}^{2} \right],\\
\label{dotrho}
\dot{\rho_\gamma}=-3H(\rho_\gamma+p_\gamma),\\
\label{ddotphi}
 \ddot{\phi} + 3H\dot{\phi}-
\frac{\dot{\theta}^{2}}{f^2}\phi +
V'(\phi )= 0,\\
\label{ddottheta}
 \ddot{\theta} + \left( 3H + 2
\frac{\dot{\phi}}{\phi}\right) \dot{\theta}+\frac{f^2}{\phi^2}
U'(\theta) = 0 ,
\end{eqnarray}
where $H$ is the Hubble parameter, dot and prime denote
derivatives with respect to $t$ and $\phi$ (or $\theta$)
respectively. We define
$\rho_{\phi}=\frac{1}{2}\dot{\phi}^{2}+V(\phi)$ and
$\rho_{\theta}=\frac{1}{2}
\frac{\phi^2}{f^2}\dot{\theta}^{2}+U(\theta)$.

\section{Phase space}
In this section, we investigate the global structure of the
dynamical system via phase plane analysis and compute the
cosmological evolution by numerical analysis. Introduce the
following dimensionless variables:

\begin{eqnarray}
x_1=\frac{\kappa}{\sqrt{6}H}\dot{\phi},\quad
y_1=\frac{\kappa\sqrt{V(\phi)}}{\sqrt{3}H}, \nonumber\\
x_2=\frac{\phi}{f}\frac{\kappa}{\sqrt{6}H}\dot{\theta},\quad
y_2=\frac{\kappa\sqrt{U(\theta)}}{\sqrt{3}H},\nonumber\\
\lambda_1=-\frac{V'(\phi)}{\kappa V(\phi)},\quad
\Gamma_1=\frac{V(\phi)V''(\phi)}{V'^2(\phi)},\nonumber\\
\lambda_2=-\frac{U'(\theta)}{\kappa U(\theta)},\quad
\Gamma_2=\frac{U(\theta)U''(\theta)}{U'^2(\theta)},\nonumber\\
z=\frac{\sqrt 6}{\kappa \phi},\quad N=\ln a,
\end{eqnarray}
and dimensionless constant
\begin{equation}
\xi=f \kappa.
\end{equation}
Note that $f = {\cal O} (M_{Pl})$, $\kappa=\sqrt{8 \pi G}$,
$G=m_{Pl}^{-2}$ (natural units) and $m_{Pl}=\sqrt{8 \pi} M_{Pl}$,
so we could have $\xi={\cal O} (1)$. It is useful when we do
numerical calculation.

Now, the equations (\ref{H2})-(\ref{ddottheta}) become the
following system:
\begin{eqnarray}\label{auto1}\fl
\frac{dx_1}{dN}=\frac{3}{2} x_1 \left
[\gamma(1-x_1^2-x_2^2-y_1^2-y_2^2)+2(x_1^2+x_2^2) \right
]-3x_1+x_2^2 z+\sqrt{\frac{3}{2}} \lambda_1 y_1^2,\nonumber\\\fl
\frac{dx_2}{dN}=\frac{3}{2} x_2 \left
[\gamma(1-x_1^2-x_2^2-y_1^2-y_2^2)+2(x_1^2+x_2^2) \right
]-3x_2-x_1 x_2 z+\frac{1}{2}\lambda_2 \xi y_2^2 z,\nonumber\\\fl
\frac{dy_1}{dN}=\frac{3}{2} y_1 \left
[\gamma(1-x_1^2-x_2^2-y_1^2-y_2^2)+2(x_1^2+x_2^2) \right ]
-\sqrt{\frac{3}{2}} \lambda_1 x_1 y_1,\nonumber\\\fl
\frac{dy_2}{dN}=\frac{3}{2} y_2 \left
[\gamma(1-x_1^2-x_2^2-y_1^2-y_2^2)+2(x_1^2+x_2^2) \right
]-\frac{1}{2}\lambda_2 \xi x_2 y_2 z,\\\fl \frac{dz}{dN}=-x_1
z^2,\nonumber\\\fl \frac{d\lambda_1}{dN} =-\sqrt{6} \lambda_1^2
x_1 (\Gamma_1-1),\nonumber\\\fl
\frac{d\lambda_2}{dN}
=-\lambda_2^2 \xi z x_2 (\Gamma_2-1).\nonumber
\end{eqnarray}
Also, we have a constraint equation
\begin{equation}\label{constraint}
\Omega_{\phi}+\Omega_{\theta}+\frac{\kappa^2\rho_\gamma}{3H^2}=1,
\end{equation}

\noindent where
\begin{eqnarray}\label{define}
\Omega_\phi\equiv{\kappa^2 \rho_\phi \over 3 H^2}=x_1^2+y_1^2,\\
\Omega_\theta\equiv{\kappa^2 \rho_\theta \over 3 H^2}=x_2^2+y_2^2.
\end{eqnarray}
The equation of state for the complex scalar field could be
expressed in terms of the new variables as

\begin{equation}\label{equaofstate}
 w_\Phi=\frac{x_1^2+x_2^2-y_1^2-y_2^2}{x_1^2+x_2^2+y_1^2+y_2^2}.
\end{equation}

According to the definitions, the parameters $\lambda_1$,
$\Gamma_1$, $\lambda_2$, $\Gamma_2$ become constants and are equal
to $\lambda_{1(0)}$, $1$, $\lambda_{2(0)}$ and $1$, respectively.
Under the circumstance, the equations (\ref{auto1}) constitute an
autonomous system as follows,
\begin{eqnarray}\label{auto2}\fl
\frac{dx_1}{dN}=\frac{3}{2} x_1 \left
[\gamma(1-x_1^2-x_2^2-y_1^2-y_2^2)+2(x_1^2+x_2^2) \right
]-3x_1+x_2^2 z+\sqrt{\frac{3}{2}} \lambda_{1(0)}
y_1^2,\nonumber\\\fl
\frac{dx_2}{dN}=\frac{3}{2} x_2 \left
[\gamma(1-x_1^2-x_2^2-y_1^2-y_2^2)+2(x_1^2+x_2^2) \right
]-3x_2-x_1 x_2 z+\frac{1}{2}\lambda_{2(0)} \xi y_2^2 z,\nonumber\\
\fl\frac{dy_1}{dN}=\frac{3}{2} y_1 \left
[\gamma(1-x_1^2-x_2^2-y_1^2-y_2^2)+2(x_1^2+x_2^2) \right ]
-\sqrt{\frac{3}{2}} \lambda_{1(0)} x_1 y_1,\\
\fl\frac{dy_2}{dN}=\frac{3}{2} y_2 \left
[\gamma(1-x_1^2-x_2^2-y_1^2-y_2^2)+2(x_1^2+x_2^2) \right
]-\frac{1}{2}\lambda_{2(0)} \xi x_2 y_2 z,\nonumber\\
\fl\frac{dz}{dN}=-x_1 z^2.\nonumber
\end{eqnarray}

When barotropic matter is under consideration or the equations of
motion are too difficult to solve analytically, phase space
methods become particularly useful, because numerical solutions
with random initial conditions usually do not expose all the
interesting properties. In table 1, we list the critical points
and the cosmological parameters there. To gain some insight into
the property of the critical points, similarly as in \cite{hao},
we investigate the stability of these critical points. For the
critical points listed in table 1, we find the eigenvalues of the
linear perturbation matrix, see table 2. For stability we require
the all 5 eigenvalues to be negative.

\begin{figure}[h]
\begin{center}
\epsfig{file=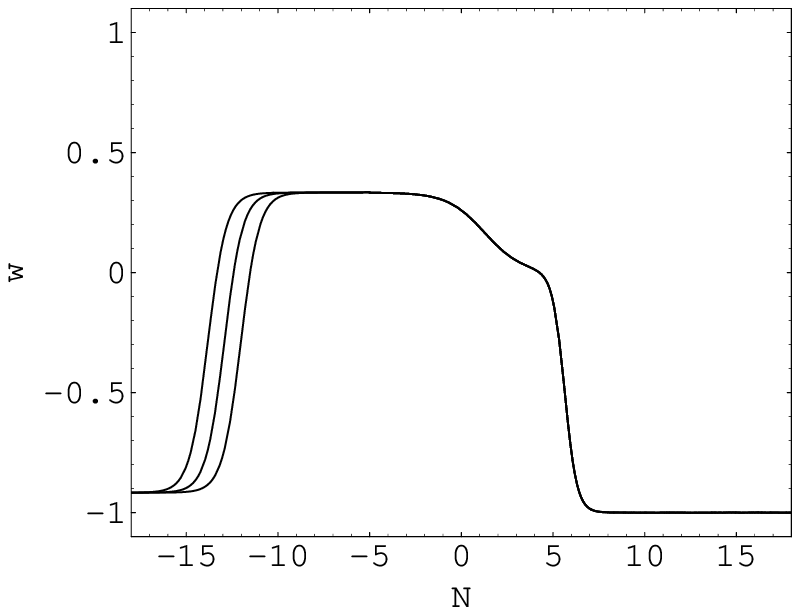,height=2.1in,width=2.6in} \caption{Evolution
of the equation of state at the presence of radiation and matter.
The curves correspond to the different initial conditions.
}\label{fig1} \epsfig{file=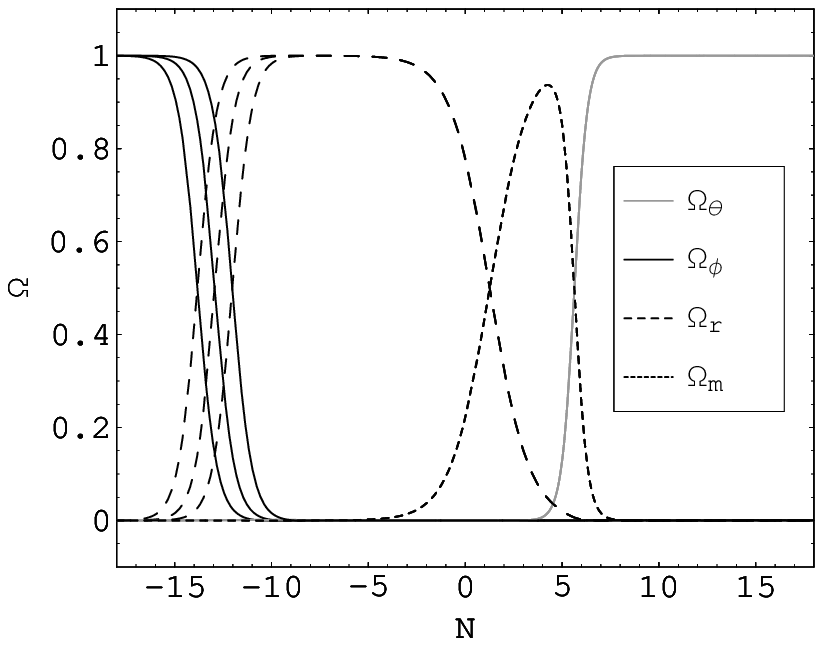,height=2.1in,width=2.6in}
\caption{Evolution of cosmic density parameter of quintessence
energy density $\Omega_\theta$, inflaton energy density
$\Omega_\phi$, radiation energy density $\Omega_r$ and matter
energy density $\Omega_m$. The curves correspond to the different
initial conditions.}\label{fig2}
\end{center}
\end{figure}

The corresponding conclusions of numerical calculation are shown
in Fig.\ref{fig1} and Fig.\ref{fig2}. This numerical solutions
describe heteroclinic orbits, which interpolate between an
unstable critical point (case i) and a late-time de Sitter
attractor (case iii). Therefore, this model is more natural to
explain the two stages of acceleration. A point worth emphasizing
is that the ordinary matter (radiation and dust) affect the
evolution of the scalar field via their contribution to the
general expansion of the Universe because of the couples can be
neglected in the model. In Fig.\ref{fig1}, the behavior of the
equation of state parameter is shown. At initial time, $w\simeq-1$
(we choose $\lambda_{1(0)}=0.5$, then $w\simeq 0.916$) corresponds
the inflationary phase, then it increases and becomes positive.
After arriving at the value $1/3$, the Universe comes to the
radiation dominated epoch and $w$ stays on a broad platform. Next,
$w$ drops to zero and stays on a narrow platform. Finally, $w$
drops below zero and approaches to $-1$, which corresponds second
stage of accelerated expansion. The evolution of cosmic density
parameters are shown in Fig.\ref{fig2}. The phase part
contribution $\Omega_\theta$ stays at $\Omega=0$ at the very first
moments of the Universe and becomes $1$ in late-time, which plays
the role of quintessence field. On the other hand, the amplitude
part contribution $\Omega_\phi$ plays the role of inflaton. During
the whole evolution, the radiation energy density and the dust
matter energy density become dominate respectively. Therefore, we
see that the constraints arising from cosmological nucleosynthesis
and structure formation are satisfied.

\begin{center}
\begin{table}[b]
\begin{tabular}{ c c c c c c c c}
  \hline
   Case & Critical points
  ($x_1$, $x_2$, $y_1$, $y_2$, z) &  $\Omega_\phi$ & $\Omega_\theta$ & $w$ \\
\hline (\romannumeral1) & $\sqrt{\frac{\lambda_{1(0)}^2}{6}}$,
$0$, $\sqrt{1-\frac{\lambda_{1(0)}^2}{6}}$, $0$, $0$ & $1$ &
$0$ & $-1+\frac{\lambda_{1(0)}^2}{3}$ \\
(\romannumeral2) & $\sqrt{\frac{3}{2}} \frac{\gamma}{\lambda_1}$,
$0$, $\sqrt{\frac{3}{2}} \sqrt{\frac{2 \gamma -
\gamma^2}{\lambda_{1(0)}^2}}$,
$0$, $0$ & $\frac{3 \gamma}{\lambda_{1(0)}^2}$  & $1$ & $\gamma -1$ \\
(\romannumeral3) & $0$, $0$, $1$, $0$, $0$ & $0$ & $1$ & $-1$ \\
(\romannumeral4) & $\pm 1$, $0$, $0$, $0$, $0$ & $1$ & $0$ & $1$ \\
(\romannumeral5) & $0$, $\pm 1$, $0$, $0$, $0$ & $0$ & $1$ & $1$ \\
(\romannumeral6) & $0$, $0$, $0$, $0$, any  & -- & -- & --\\
\hline
\end{tabular}
\caption{The properties of the critical points.}
\end{table}\label{table1}
\end{center}

\begin{center}
\begin{table}[t]
\begin{tabular}{ c c c c c c c c}
  \hline
   case & corresponding eigenvalues & stability\\
\hline (\romannumeral1) & $0$, $\frac{\lambda_{1(0)}^2}{2}$,
$-3+\frac{\lambda_{1(0)}^2}{2}$,
$-3+\frac{\lambda_{1(0)}^2}{2}$, $-3 \gamma+\lambda_{1(0)}^2$ & unstable\\
(\romannumeral2) & $0$, $\frac{3 \gamma}{2}$, $\frac{3}{2}
(\gamma-2)$,\\ &
$\frac{3\{\lambda_{1(0)}^2(\gamma-2)-\sqrt{\lambda_{1(0)}^2(\gamma-2)
[\lambda_{1(0)}^2(9\gamma-2)-24\gamma^2]}\}}{4\lambda_{1(0)}^2}$,\\
&
$\frac{3\{\lambda_{1(0)}^2(\gamma-2)+\sqrt{\lambda_{1(0)}^2(\gamma-2)
[\lambda_{1(0)}^2(9\gamma-2)-24\gamma^2]}\}}{4\lambda_{1(0)}^2}$ & unstable\\
(\romannumeral3) & $-3$, $0$, $0$, $0$, $-3 \gamma$ & stable\\
(\romannumeral4) & $3$, $0$, $0$, $6-3 \gamma$,
$3\mp\sqrt{\frac{3}{2}} \lambda_{1(0)}$
& unstable \\
(\romannumeral5) & $3$, $3$, $0$, $0$, $6-3 \gamma$ & unstable\\
(\romannumeral6) & $0$, $\frac{3 \gamma}{2}$, $\frac{3
\gamma}{2}$,
$-3+\frac{3 \gamma}{2}$, $-3+\frac{3 \gamma}{2}$ & unstable\\
\hline
\end{tabular}
\caption{The eigenvalues of the critical points.}
\end{table}\label{table2}
\end{center}

\section{Conclusions}

In this paper, we have discussed the cosmological implication of a
complex scalar model of quintessence inflation with a barotropic
fluid. Using the numerical calculation for this model, we show
that the phase contribution is negligible at the early epoch of
the universe while it becomes dominate with the time evolving and
the evolution of the amplitude part is the opposite, which is a
viable way to unify the description of the inflationary stage and
the current accelerated expansion. Analysis to the dynamical
evolution of the complex scalar model indicates that it admits a
late time attractor solution, at which the field behaves as a
cosmological constant. Obviously, attractor and heteroclinic orbit
are both alleviate the fine tuning problem.

\section*{Acknowledgment}
This work is supported by National Natural Science Foundation of
China under Grant No.10473007 and the Foundation from Science and
Technology Committee of Shanghai under Grant No.02QA14033.


\section*{References}
\begin{thebibliography}{99}
\bibitem{CMB} C. L. Bennett \textit{et al.}, Astrophys. J. Suppl. \textbf{148} (2003) 1.
\bibitem{SNIa} J. L. Tonry \textit{et al.}, Astrophys. J. \textbf{594} (2003) 1.
\bibitem{SDSS} M. Tegmark \textit{et al.}, Astrophys. J. \textbf{606} (2004) 702.
\bibitem{SingleField} P. J. E. Peebles and B. Ratra, {Rev. Mod. Phys}. \textbf{75} (2003) 599;
\\ T. Padmanabhan, {Phys. Rep}. \textbf{380} (2003) 235.
\bibitem{NCoupledField} J. G. Hao and X. Z. Li, Class. Quant. Grav. \textbf{21} (2004) 4771.
 \\X. Z. Li, J. G. Hao and D. J. Liu, Int.J.Mod.Phys. \textbf{A18} (2003) 5921;
\\ X. Z. Li and J. G. Hao, Phys. Rev. \textbf{D69} (2004) 107303;
\\ J. G. Hao and X. Z. Li, Class. Quant. Grav. \textbf{21} (2004)
4771.

\bibitem{Caldwell} R. R. Caldwell, Phys. Lett. \textbf{B545} (2002) 23.

\bibitem{BIL} J. G. Hao and X. Z. Li, Phys. Rev. \textbf{D68} (2003)
043501;\\ D. J. Liu and X. Z. Li, Phys. Rev. \textbf{D68} (2003)
067301;\\ J. G. Hao and X. Z. Li, Phys. Rev. \textbf{D68} (2003)
087301.

\bibitem{k-essence} C. Armendariz-Picon, V. Mukhanov and P. J.
Steinhardt, Phys. Rev. \textbf{D63} (2001) 103510.

\bibitem{GCG} M. C. Bento, O. Bertolami and A. A. Sen, Phys.
Rev. \textbf{D67} (2003) 063003;
 \\J. G. Hao and X. Z. Li, Phys. Lett. \textbf{B606} (2005) 7;
 \\D. J. Liu and X. Z. Li, Chin.Phys.Lett. \textbf{22} (2005) 265.
\bibitem {Kolb&Turner} E. W. Kolb and M. S. Turner, \textit{The Early Universe} (Addison-wesley, Reading, MA,
1990); \\A. D. Linde, \textit{Particle Physics and Inflationary
Cosmology} (Harward Academic, Chur, Switzerland, 1990).

\bibitem {Liddle&Lyth} A. R. Liddle and D. H. Lyth, \textit{Cosmological
Inflation and Large-Scale Structure} (Cambridge University Press,
2000).

\bibitem{QuintInf} P. J. E. Peebles and A. Vilenkin, Phys. Rev. \textbf{D59} (1999)
063505.


\bibitem{complexscalar} R. Rosenfeld and J. A. Frieman, astro-ph/0504191;
\\ J. A. Gu and W. Y. P. Hwang, Phys. Lett. \textbf{B517} (2001) 1.

\bibitem{hao} J. G. Hao and X. Z. Li, Phys. Rev.
\textbf{D70} (2004) 043529;\\ D. J. Liu and X. Z. Li, Phys. Lett.
(2005) \textbf{B611} 8

\bibitem{heteroclinic} X. Z. Li, Y. B. Zhao and C. B. Sun,
astro-ph/0508019, Class. Quant. Grav. in press.


\end{thebibliography}
\end{document}